\documentclass[twocolumn,amsmath,amssymb, aps, prl, groupedaddress, showpacs, showkeys]{revtex4-1}

\usepackage[english]{babel}
\usepackage{graphicx}
\usepackage{color}

\begin{document}

\title{Curling Liquid Crystal Microswimmers: a cascade of spontaneous symmetry breaking}

\author{Carsten Kr\"{u}ger}
\affiliation{Max Planck Institute for Dynamics and Self-Organization (MPIDS), Am Fa\ss{}berg 17, 37077 G\"{o}ttingen, Germany}

\author{Gunnar Kl\"{o}s}
\affiliation{Max Planck Institute for Dynamics and Self-Organization (MPIDS), Am Fa\ss{}berg 17, 37077 G\"{o}ttingen, Germany}

\author{Christian Bahr}
\affiliation{Max Planck Institute for Dynamics and Self-Organization (MPIDS), Am Fa\ss{}berg 17, 37077 G\"{o}ttingen, Germany}

\author{Corinna C. Maass}
\affiliation{Max Planck Institute for Dynamics and Self-Organization (MPIDS), Am Fa\ss{}berg 17, 37077 G\"{o}ttingen, Germany}

\bibliographystyle{apsrev4-1}

\begin{abstract}
We report curling self-propulsion in aqueous emulsions of common mesogenic compounds. Nematic liquid crystal droplets self-propel in a surfactant solution with concentrations above the critical micelle concentration while undergoing micellar solubilization \cite{Herminghaus2014}. We analyzed trajectories both in a Hele-Shaw geometry and in a 3D setup at variable buoyancy. The coupling between the nematic director field and the convective flow inside the droplet leads to a second symmetry breaking which gives rise to curling motion in 2D. This is demonstrated through a reversible transition to non-helical persistent swimming by heating to the isotropic phase. Furthermore, auto-chemotaxis can spontaneously break the inversion symmetry, leading to helical trajectories.
\end{abstract}%

\maketitle

Artificial self-propelled systems have gained attention as small scale model systems for simulating biological equivalents, either for single particles or large scale collective behaviour. In this framework, many different schemes for experimental self-propelling swimmers have been developed, mimicking various features of biological systems. The experiments can be categorized either as surfers needing direct contact with an interface \cite{DosSantos1995,Ban2008,Chen2009,Palacci2013}, or swimmers which self-propel in the bulk. Prime examples for the latter case are Janus particles \cite{Howse2007,Volpe2011,Buttinoni2013,Kuemmel2013,Hagen2014} and active emulsions of droplets in surfactant solutions \cite{Toyota2006,Hanczyc2007,Thutupalli2011,Banno2012,Ban2013,Banno2013,Peddireddy2012,Herminghaus2014,Izri2014, Maass2016}. So far, all swimmer systems with spherical or polar symmetry show active Brownian behavior, with trajectories that are ballistic over short distance and diffusive over long timescales \cite{Howse2007,Thutupalli2011,Volpe2011,Izri2014}. In contrast, helical or circular trajectories have been reported in bioflagellates close to surfaces \cite{Lauga2006,Celli2009,Hu2015}, as well as in bacterial and artificial swimmers with at least twofold structural asymmetry, both in experiments and simulations~\cite{Crenshaw1996,Teeffelen2008,LedesmaAguilar2012,Kuemmel2013,Hagen2014,Hagen2015}.

In this paper, we report curling motion in two, as well as helical trajectories in three dimensions, for active emulsions.  Our system that is spherically symmetric at rest, consisting of nematic liquid crystal droplets in an aqueous surfactant solution \cite{Peddireddy2012,Herminghaus2014}. Rotational torques are generated by the interplay of surface flow and nematic order, such that the curling motion can be switched off by heating to the isotropic state and the droplet reverts to persistent swimming. Since fundamental parameters and forces can be tuned reproducibly, such emulsions are well suited for comparison to current numerical studies \cite{Wittkowski2012}.

We study droplets of the calamitic liquid crystal 4-pentyl-4$^\prime$-cyanobiphenyl (5CB), which is nematic at room temperature, in an aqueous solution of the ionic surfactant tetra\-decyltrimethylammonium bromide (TTAB). If the TTAB concentration exceeds the critical micelle concentration, $c_\text{TTAB} > \text{CMC}\approx 0.13$\,wt\%, the droplets slowly dissolve by micellar solubilization, with the droplet radius decreasing linearly with time \cite{Peddireddy2012}. For TTAB concentrations above $c_\text{TTAB}\approx 5$\,wt\%, the droplets self propel with typical speeds between 5 and 25\,$\mu$m\,s$^{-1}$ \cite{Herminghaus2014}. Shrinking rate and speed increase with increasing $c_\text{TTAB}$, with some saturation effects for $c_\text{TTAB}\gtrsim 17.5$\,wt\%. The droplet propulsion is caused by a self-sustained polar gradient in the surfactant coverage on the interface between the droplet and the surrounding continuous phase. This gradient in surface energy leads to Marangoni stresses along the interface, which in turn generate flows propelling the droplet. An advection-diffusion model considering the dynamics of surfactant monomers, empty and filled micelles as well as liquid crystal molecules, is able to explain this dynamic instability qualitatively \cite{Herminghaus2014,Maass2016}. In addition to such hydrodynamic propulsion forces, the droplets are also sensitive to gradients in empty micelle concentration. Since each droplet leaves a trail of filled micelles, this sensitivity induces them to avoid each other's as well as their own trajectories. Consequently, they show negative autochemotaxis~\cite{Kranz2015} on time scales faster than micellar diffusion. However, this autochemotactic repulsion is not strong enough to entirely suppress self crossing trajectories.

We will first analyze trajectories from quasi-2D Hele-Shaw cells. Here, we varied $c_\text{TTAB}$ between 7.5\,wt\% and 15\,wt\%, which is well above the CMC and inside the range of droplet propulsion. Droplets of 5CB with an initial diameter of 50 $\mu$m were injected into a quasi two dimensional reservoir of area $10 \times 6$ mm and height $h\approx 50$ $\mu$m, prefilled with TTAB solution. Trajectories were recorded at 4\,fps under a bright field microscope at $2\times$ magnification. The droplet positions were extracted from the video data via digital background correction, threshold binarization and contour analysis. Trajectories were reconstructed using a frame by frame nearest neighbor analysis. 

\begin{figure}[t]
\includegraphics[width=\columnwidth]{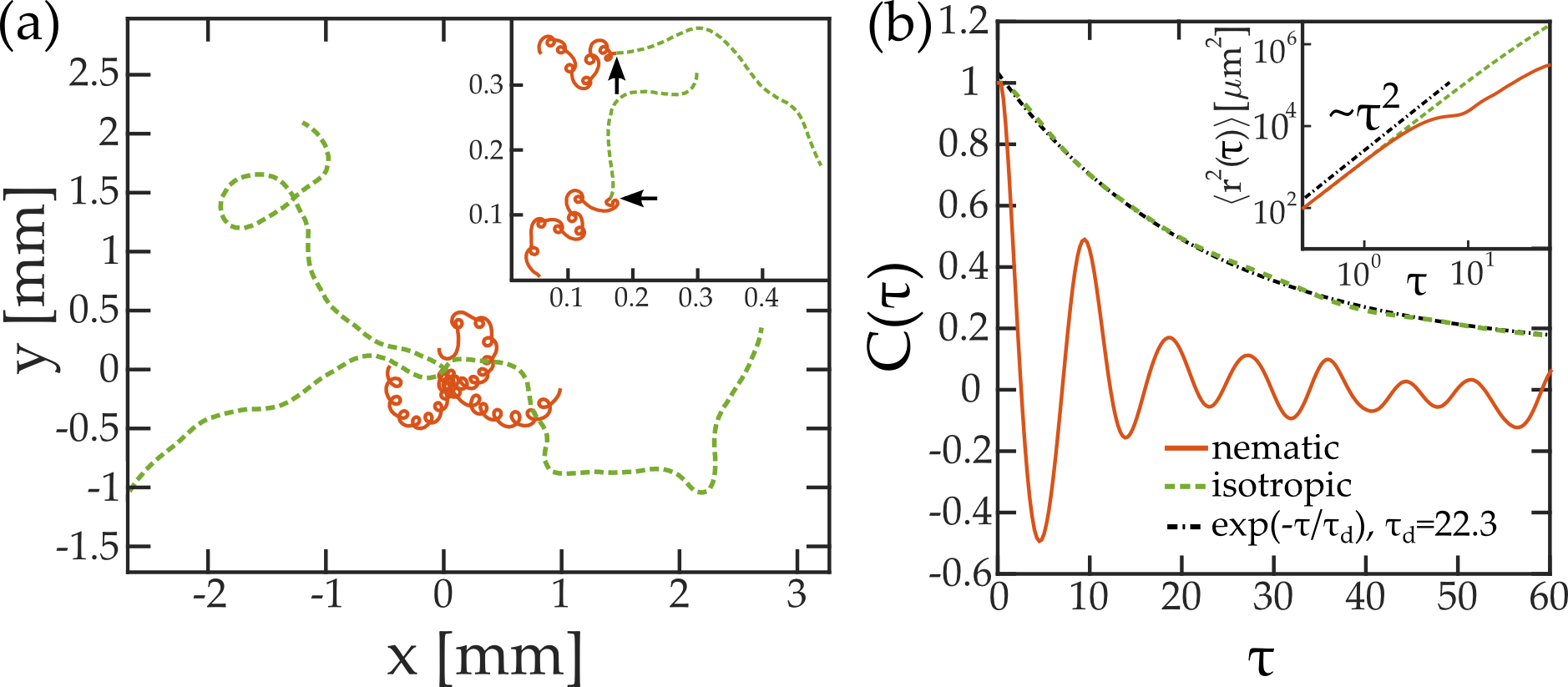}
\caption{\label{nem_iso}(Color online) (a) Example trajectories of 50\,$\mu$m droplets at $c_\text{TTAB} =7.5$\,wt\% over the course of 90\,s at $34\,^\circ$C (nematic, solid line) and $37\,^\circ$C (isotropic, dashed). Above $T_\text{NI}$, the curling is suppressed. Inset: trajectories of 100\,s duration with a temperature ramp across the nematic-to-isotropic transition. The black arrows emphasize the transition. (b):  Comparison of angular autocorrelation $C(\tau)$ in the rescaled time frame $\tau$. The isotropic $C(\tau)$ fits $\exp(-\tau/\tau_d)$ with a decay time $\tau_d = 22.3$ (dash-dotted, black). Inset: Comparison of MSD $\langle \Delta r^2(\tau) \rangle$.%
}
\end{figure}

As shown in Fig.~\ref{nem_iso}(a), the droplet trajectories show a pronounced curling when observed at temperatures below the nematic to isotropic transition (solid to dashed lines), $T_\text{NI} \approx 35\,^\circ$C. When heated above the transition, the curling motion ceases immediately [see inset in Fig.~\ref{nem_iso}(a)], and the trajectory reverts to persistent swimming, which is nearly ballistic on the time scale of observation. We should note that no chiral component is present in our system, and that clockwise and counterclockwise rotations are observed with equal probability. 
Self-propulsion was quantified via the mean squared displacement (MSD),
\begin{equation}
\langle \Delta r^2(t) \rangle  = \langle (\textbf{r}(t_0 + t) - \textbf{r}(t_0))^2	\rangle_{t_0},
\label{eq:msd}
\end{equation}
and the angular autocorrelation function $C(t)$ of the trajectories,
\begin{equation}
C(t) = \left\langle\frac{ \textbf{v}(t_0+t) \cdot \textbf{v}(t_0)}{ |\textbf{v}(t_0+t)||\textbf{v}(t_0)|} \right\rangle_{t_0}.
\label{eq:angcoll}
\end{equation}
As the droplets dissolve continuously, neither speed nor diameter are constant over time, which has to be taken into account for long time correlations. A simple method to derive a natural, unit free progressive scale $\tau$ for the system uses the accumulated trajectory length in units of the current diameter $D(t)$ by integrating over the instantaneous speed $v(t)$:

\begin{equation}
\tau = \int_0^t {\frac{v(t^\prime)}{D(t^\prime)} dt^\prime}.
\label{eq:rescaling}
\end{equation}

\begin{figure}[t]
\includegraphics[width=\columnwidth]{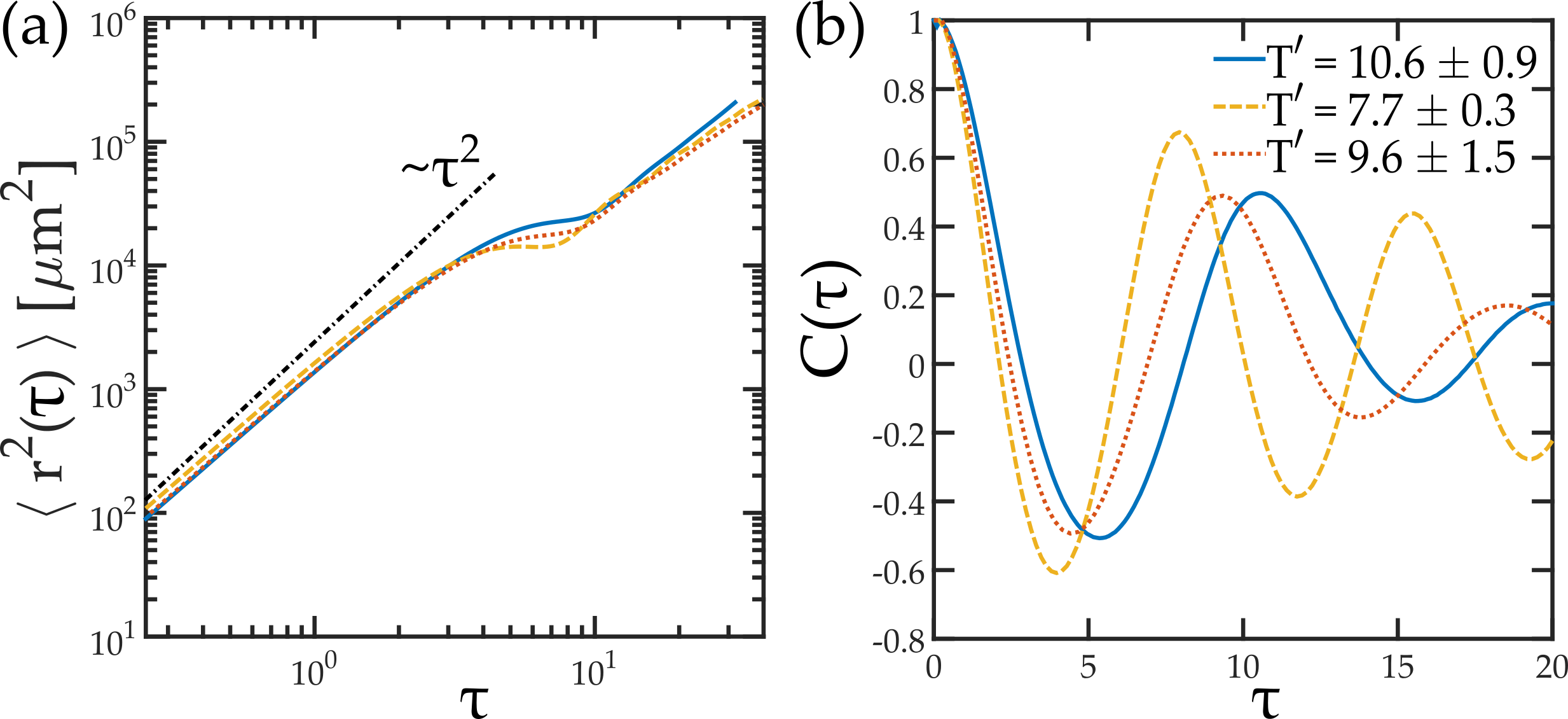}
\caption{\label{msd_corr_time}(Color online)  MSD and angular autocorrelation $C(\tau)$ for experiments in the nematic state with $c_\text{TTAB} = 7.5$\,wt\%  at 22 (solid), 30 (dashed) and $34\,^\circ$C (dotted) in the rescaled time frame $\tau$. We see data collapse by rescaling for MSD and $C(\tau)$ up to one rotation period $T'$.%
}
\end{figure}

We have plotted $\langle \Delta r^2(\tau) \rangle$ and $C(\tau)$ in Fig.~\ref{msd_corr_time} for a series of experiments taken at constant $c_\text{TTAB}$ while increasing the temperature in the range below $T_\text{NI}$. This leads to an increase in swimming speed while retaining the curling motion. Each data set contains between 12--16 trajectories similar to those plotted in Fig.~\ref{nem_iso}(a), recorded each over a time of 4-5 minutes. Individual droplet sizes vary between 30 and 45\,$\mu$m. Rescaled MSDs are plotted in Fig.~\ref{msd_corr_time}(a). The characteristic MSD for this type of swimmer is ballistic, $\propto \tau^2$, for short times up to the rotation period, where we see a dip caused by the trajectory bending back on itself while curling. On longer time scales the MSD recovers towards a slightly subballistic behavior, as can be expected from the high trajectory persistence beyond the superimposed curling. We do not see a transition to diffusive behavior, $\propto \tau$, as reported for other active swimmer systems \cite{Howse2007,Volpe2011}, as we are constrained in terms of droplet lifetime and cell size. The rescaling to the $\tau$ time scale yields a good collapse for ballistic, rotation and recovery regimes, with the signature dip at $\tau\approx 10$.

The angular autocorrelation is expected to be constant for ideal ballistic swimmers, $C(\tau)=1$, and exponentially decaying in the case of diffusive behavior with persistence length $L_p$, $C(\tau)=e^{-\tau v/L_p}$. In our system the curling causes a superimposed characteristic oscillation $\propto \cos 2\pi T^\prime$ with the rotation period $T^\prime$ of the swimmer. Rescaling to the new time scale is successful up to one period $T^\prime \approx 10$ [Fig.~\ref{msd_corr_time}(b)] consistent with the dip in the rescaled MSD.  

There is strong evidence that nematic ordering is necessary for the curling instability. In Fig.~\ref{nem_iso}(a) we show trajectories for the same surfactant concentration $c_\text{TTAB} = 7.5$\,wt\% at temperatures $T=34\,^\circ$C and $37\,^\circ$C, i.~e.~slightly below and above $T_\text{NI}$. The curling is entirely suppressed for $T>T_\text{NI}$. As soon as the transition temperature is reached, the curling motion changes to a smooth trajectory with exponentially decaying angular autocorrelation [Fig.~\ref{nem_iso}(b)], corresponding to a quite large persistence length ($L_p\approx 20\,D$). The signature dip in $\langle \Delta r^2 (\tau) \rangle$ and oscillation in $C(\tau)$ observed for the curling case are fully suppressed for isotropic droplets.

\begin{figure}[b]
\includegraphics[width=\columnwidth]{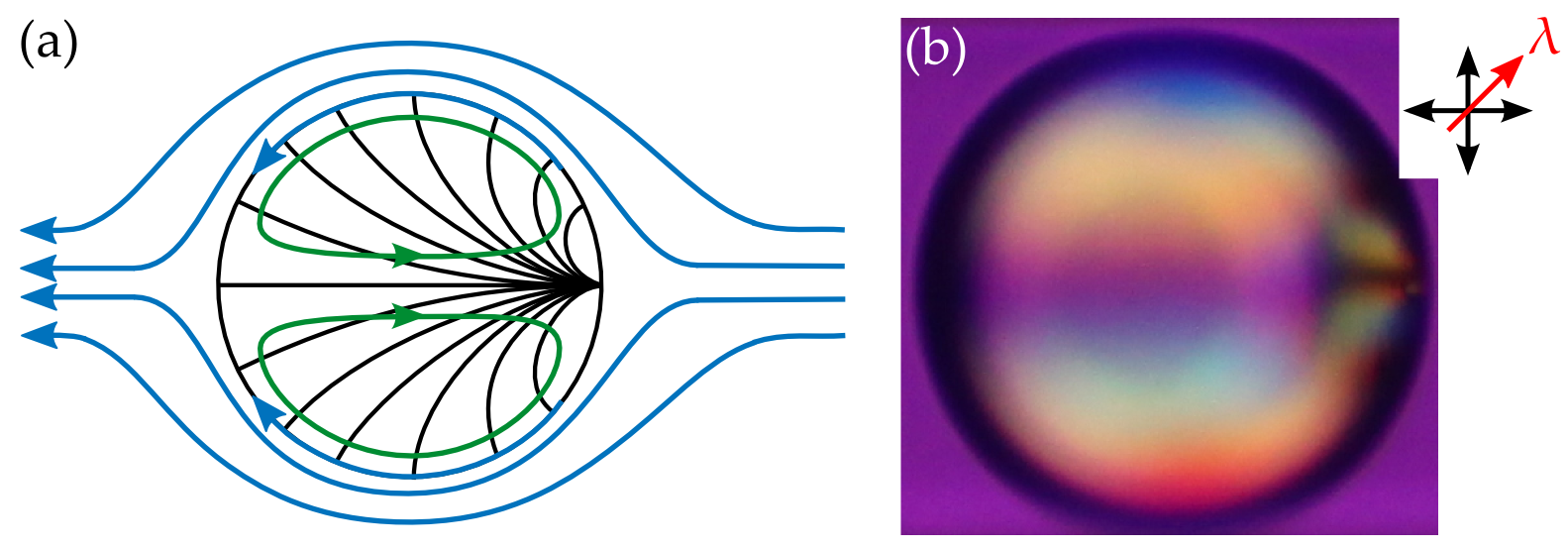}
\caption{\label{f_lc_drop} (Color online) Self-propelling nematic droplet. (a) Schematic flow field (arrows) inside and outside of the droplet and distorted nematic director field (black lines). The Marangoni flow (arrows on the interface) induced by the inhomogeneous surfactant coverage leads to internal convection and self-propelled motion. (b) A 5CB droplet ($D\approx 50\,\mu$m), moving from left to right in a straight capillary, imaged between crossed polarizers with an added $\lambda$ retardation plate. Schematic arrows denote polarizer and $\lambda$ plate orientations.%
}
\end{figure}

We found the physical mechanism underlying the curling motion by studying the nematic structure of a moving droplet observed between crossed polarizers. The surfactant layer at the surface of the droplet causes a homeotropic anchoring of the nematic director at the droplet interface, resulting in a topological point defect (radial hedgehog) in the center of a resting droplet \cite{Candau1973}. The observed birefringence patterns show that the internal convection of a self-propelling droplet [Fig.~\ref{f_lc_drop}(a)] advects the point defect towards the leading edge of the droplet [Fig.~\ref{f_lc_drop}(b)]. 

\begin{figure*}
\includegraphics[width=.8\textwidth]{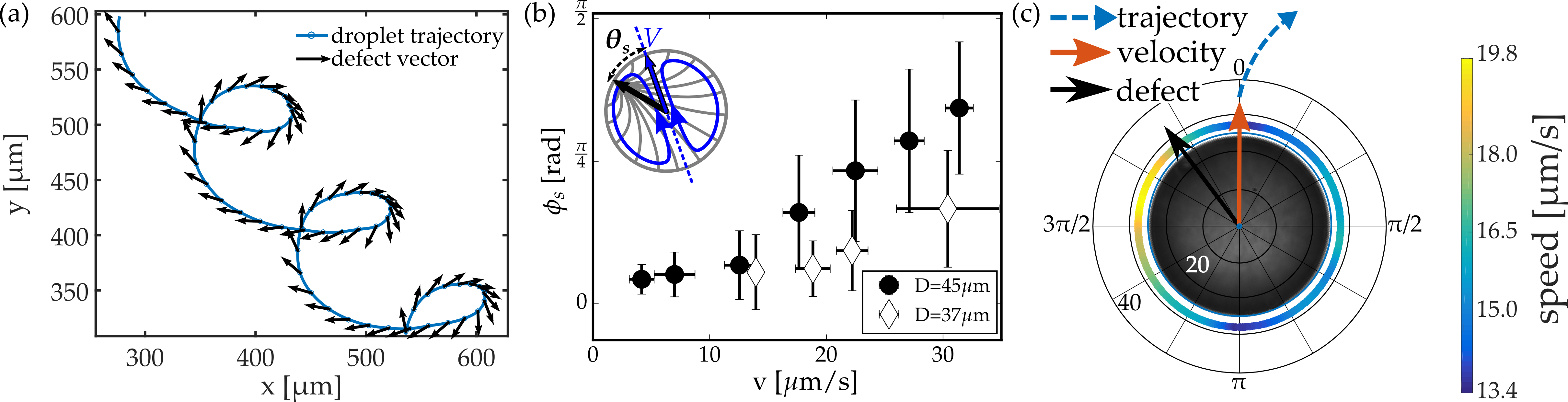}
\caption{\label{angle_surfvel} (Color online) (a) Curling trajectory of a 50\,$\mu$m droplet with defect vectors (arrows) and swimmer trajectory (solid line). (b) Averaged phase shift $\theta_s$ between velocity and defect vs. droplet speed for two droplet sizes. (c) Tangential velocity $\approx$ 4\,$\mu$m away from the droplet interface. The flow is increased around the defect location.%
} 
\end{figure*}

The micrograph shown in Fig.~\ref{f_lc_drop}(b) was obtained for a droplet in a straight capillary with a square cross section equal in size to the droplet diameter. In this one dimensional confinement, the droplet is forced to follow the capillary axis and the defect is located at the apex of the droplet. Intriguingly, for curling motion in a two-dimensional reservoir, we always observe a finite angle $\theta_s$ between the velocity vector and the defect vector pointing from the droplet center to the defect location [Fig.~\ref{angle_surfvel}(a) and inset in Fig.~\ref{angle_surfvel}(b)]. This can be explained by the fact that the defect orientation towards the stagnation point at the droplet apex is not stable against fluctuations in the polar angle $\theta$, since the surface flow $v_s$ follows $v_s\propto \sin\theta$ to a first order \cite{Blake1971}.
Accordingly, the defect is pulled away from the apex, until this deflection is countered by elastic restoring forces caused by the deformation of the nematic director field. Since tangential stresses are continuous across the interface, but there is no elastic restoring force in the aqueous phase, we have a net tangential force pulling the outer phase towards the droplet equator, resulting in a counterrotating torque acting on the droplet.

Microscopically, due to the homeotropic anchoring, the flow of the nematic material near the droplet surface is balanced by the Miesowicz viscosity $\eta_2$ (director parallel to the velocity gradient), which is the largest of the three nematic viscosity coefficients \cite{Miesowicz1946}. Around the defect, however, the director is rotated away from the interface normal and the nematic order parameter is decreased as well. This causes the effective viscosity to decrease towards the isotropic average $\eta_\text{iso}\approx \frac13 \eta_2$ \cite{Chmielewski1986}. In consequence, the Marangoni surface flow, which drives the droplet motion, will be faster near the defect and this asymmetry adds a rotational component to the motion of the droplet. By tracking 1\,$\mu$m colloidal tracers in the continuous phase close ($\approx4\,\mu$m) to the interface we indeed find that the surface velocity increases near the defect location. The defect vector always points to the convex side of the curved trajectory following the faster surface flow [Fig.~\ref{angle_surfvel}(c)]. 

We have plotted the average angle $\theta_s$ between the defect and velocity vectors versus the average droplet speed $v$ in Fig.~\ref{angle_surfvel}(b). $\theta_s$ increases with $v$ in a nonlinear manner. Qualitatively, we see that with increasing $v$ the defect will be pulled further towards the droplet equator, where in turn the surface speed $v_s$ increases, such that we can expect a $\theta_s \propto v^2$ component for small $\theta_s$, where $v_s\propto \sin\theta \approx \theta$.  For smaller droplets, $D=37\,\mu$m, $\theta_s$ increases less with $v$, since the same topological charge applies to a smaller volume, thus requiring a higher deformation energy. A more quantitative treatment would require a precise numerical estimation of the interplay of convective flow and nematic ordering and is beyond the scope of this paper.

\begin{figure}[b]
\centering\includegraphics[width=\columnwidth]{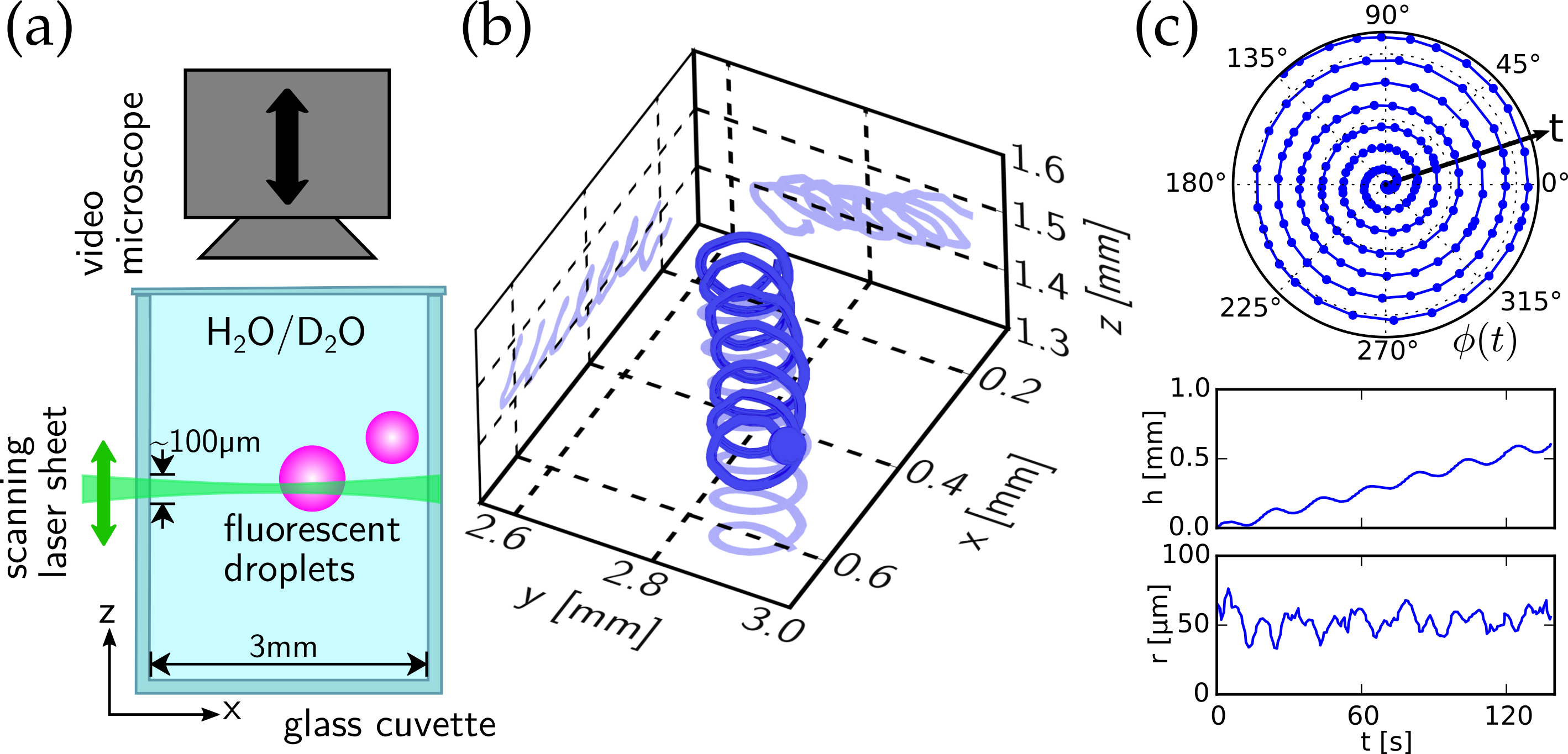}
\caption{\label{LSM}(Color online) (a) Light sheet microscope: Fluorescently dyed droplets are excited by a wide laser sheet scanning the sample in $z$. A video microscope focused on the sheet records time and $xy$ positions. (b) reconstructed helical 3D trajectory of a buoyant 50\,$\mu$m swimmer at $c_\text{TTAB}=7.5$\,wt\%. (c) azimuthal angle $\phi$, height $h$ along axis and curvature $r$ vs. time for the same trajectory.
}
\end{figure} 

We further needed to address the question how the droplets propel in an unconfined environment, since rotation can also be caused by interaction with a solid interface, as known from bioflagellates like \textit{E. coli} \cite{Lauga2006, Celli2009}. By adding heavy water, D$_2$O, to the surfactant solution, we matched the densities of the aqueous phase and the swimming droplets.  We recorded 3D trajectories with a custom built light sheet fluorescence microscope in a volume of $3 \times 3 \times 3$ mm$^3$, as shown in Fig.~\ref{LSM}, and found helices with well defined pitch $p$, azimuthal angle $\phi$ and curvature $r$ over the persistence length of the overall trajectory (cf. example in Fig.~\ref{LSM} (b), (c) with $r=55\pm4\,\mu$m, $p=100\pm20\,\mu$m, $v=17.4\pm3\,\mu$m). The experimental conditions were kept similar to the quasi 2D study, except for the addition of D$_2$O to the surfactant solution and the fluorescent dopant Nile Red to the 5CB. 

\begin{figure}[t]
\centering\includegraphics[width=\columnwidth]{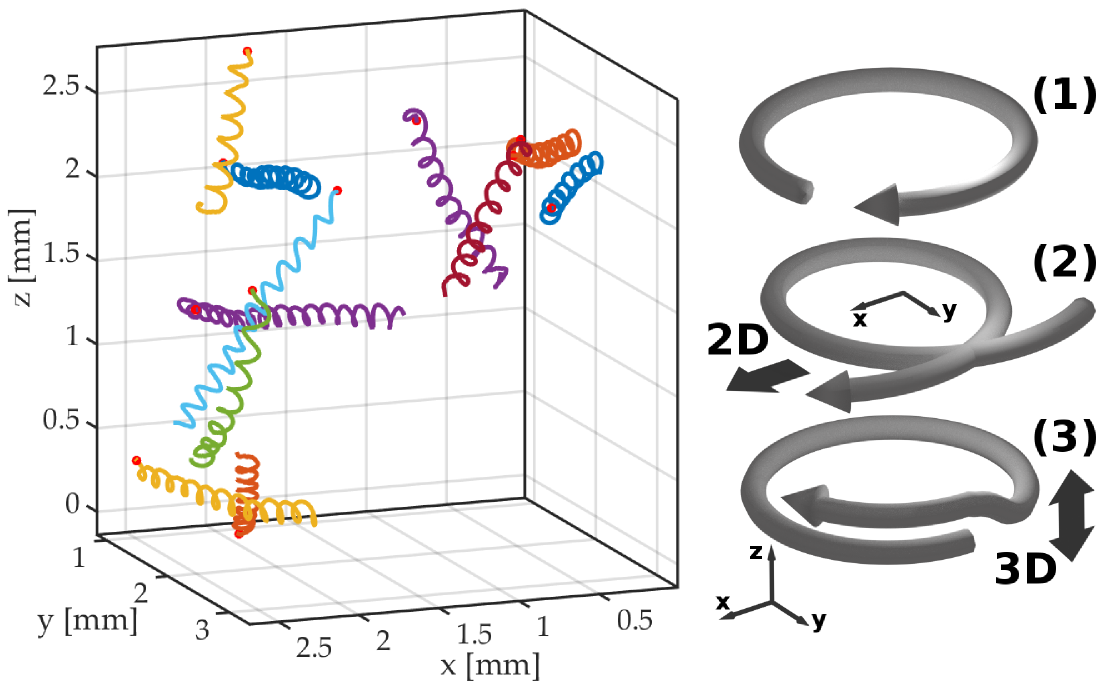}
\caption{\label{LSMstack}(Color online) Left: Trajectories of 11 droplets in a buoyancy matched aqueous TTAB solution of 5\,wt\%, at room temperature, over a time of 5 minutes, with dots marking the initial position. The trajectories show no global bias of orientation or chirality (5 right, 6 left handed). Right: Conceptual sketch of trail avoidance strategies after the initial undisturbed orbit (1). In 2D (2), the resulting autochemotactic force (black arrows) is in plane and leads to a trajectory crossing, in 3D (3) parallel or antiparallel to the orbit normal.
}
\end{figure}

This helical instability cannot be caused solely by a static deflection of the defect vector: If the force on the defect is purely polar without any azimuthal component, as assumed for our quasi 2D system, the trajectory should be circular and confined to the plane spanned by the velocity and the defect vector [see Fig.~\ref{LSMstack}, right (1)] \cite{Wittkowski2012}. However, if a swimmer circles around to meet its trajectory, it will encounter a repulsive chemotactic gradient caused by its own trail of filled surfactant micelles, in turn leading to an out of plane force component. Effectively, the swimmer is pushed into a direction perpendicular to the plane of rotation, leading to a stable helical trajectory within the time frame of persistent swimming [Fig.~\ref{LSMstack}, right (3)]. A constant external force like gravity would also lead to helical trajectories, but with the helical axis aligning with the force vector. Trajectories from an experiment with several droplets are shown in the left panel of Fig.~\ref{LSMstack}. The helix axes show no global directional bias, including gravity, $\vec{g}\parallel\vec{z}$. Of the eleven trajectories recorded, five are right handed and six left handed. This equidistribution confirms that there is no global chirality in the system, as should be expected from the symmetry of the underlying flow instability. The chemotactic repulsion is probably also the reason for the behavior in the 2D environment [Fig.~\ref{LSMstack}, right (2)], where the droplets do not move on a simple circular trajectory but show a net propagation away from regions containing filled micelles.

We have reported on active droplets exhibiting a unique two-step spontaneous symmetry breaking. The first step results in self-propelled motion while the second, which is present only when the droplets are in the nematic state, adds a rotational component to the active motion. Thus, although the droplets are radially isotropic at rest, they show curling motion in 2D and helical swimming in 3D, entirely from spontaneously emerging dynamic flow instabilities and chemotactic effects. The active mesogenic droplets provide a simple experimental model system which is especially useful with respect to numerical and analytical tractability as the interfacial geometry is spherical, thereby enabling a description in which rotation and translation are decoupled \cite{Wittkowski2012}. Moreover, the coupling between surface flow, internal convection, and nematic director field provides intriguing possibilities to control the swimming behavior. The most obvious is simply to switch off the torque causing the curling by heating to the isotropic phase, as was demonstrated in the present study. A more subtle control could be achieved by using a nematic compound possessing a smectic-$A$ phase (e.\ g., the longer homolog 8CB). Since the nematic twist and bend elastic constants show a divergent behavior when the transition to the smectic-$A$ phase is approached \cite{Sprunt1984}, the deformation of the director field by the internal convection, and hence the torque acting on the droplet, will decrease with decreasing temperature difference to the transition. Thus, quantities such as the curvature of the trajectory could be tuned over a wide range by adjusting the temperature.

Financial support from the Deutsche Forschungsgemeinschaft (Priority Programme SPP1726 ``Microswimmers'') is gratefully acknowledged.

\bibliographystyle{apsrev4-1}

\bibliography{bibliography}

\end{document}